\documentclass[prl,showpacs,twocolumn,aps,a4paper]{revtex4}
\usepackage{dcolumn}
\usepackage{amsmath}
\usepackage{graphicx}
\usepackage{latexsym}
\usepackage{amsfonts}
\usepackage{amssymb}

\DeclareGraphicsExtensions{.pdf,.gif,.jpg}

\def\a{\alpha}

\def\l{\lambda}
\def\m{\mu}

\def\s{\sigma}

\def\w{\omega}

\def\ua{\uparrow}
\def\da{\downarrow}

\def\bra{\langle}
\def\ket{\rangle}

\newcommand{\be}{\begin{equation}}
\newcommand{\ee}{\end{equation}}
\newcommand{\beq}{\begin{eqnarray}}
\newcommand{\eeq}{\end{eqnarray}}

\begin{document}


\title{Dynamical formation and 
manipulation of Majorana fermions in driven quantum wires}

\author{E. Perfetto}
\affiliation{Dipartimento di Fisica, Universit\`{a} di Roma Tor
Vergata, Via della Ricerca Scientifica 1, I-00133 Rome, Italy}

\begin{abstract}
Controlling the dynamics of Majorana fermions (MF) subject to time-varying driving 
fields is of fundamental importance for the practical realization 
of topological quantum computing.
In this work we study how it is possible to dynamically generate and maintain
the topological phase in one-dimensional superconducting nanowires 
after the temporal variation of the Hamiltonian parameters. Remarkably we show that for a
sudden quench the system can never relax towards a state exhibiting fully developed MF,
independently of the initial and final Hamiltonians. Only for sufficiently slow protocols
the system behaves adiabatically, and the topological phase can be reached.
Finally we address the crucial question of how ``adiabatic'' a protocol must
be in order to manipulate the MF inside the topological phase without deteriorating their Majorana character.

 %
 %
 %
 %
 %
 %

\end{abstract}

\maketitle

{\it Introduction.---}
The enormous potential of Majorana fermions (MF) for implementing 
decoherence-free quantum computation\cite{kitaev} is 
greatly stimulating their search in solid state systems.
Several different platforms have been proposed for realizing MF, like   
fractional quantum Hall states at filling factor $5/2$,
vortex cores of $p+ip$ superconductors\cite{ivanov}, surfaces of topological 
insulators  coupled to $s$-wave superconductors\cite{kane},
non-centrosymmetric superconductors\cite{sato}, and 
ferromagnetic Josephson junctions\cite{black}.
However, despite such intense theoretical activity,
no experimental evidence of the existence of MF in the above systems  
has been provided so far.
A promising alternative that should drastically simplify the 
realization and detection  of MF consists
in forming one-dimensional (1D) heterostructures with semiconductors and conventional
superconductors\cite{sarma,vonoppen}.
Here the $p$-wave superconductivity 
is simulated by means of spin-orbit (SO) 
interaction and the system can be driven into the
topological (T) phase (where the MF appear) by applying a suitable 
magnetic field.
In addition MF occurring in 2D and 3D networks
of these nanowires are particularly attracting, since
they can be adiabatically manipulated by using tunable gates or
by reorienting of the magnetic field\cite{vonoppen,vonoppen2}.
Very recently signatures of MF presence in InSb
and InAs nanowires
have been reported\cite{exp1,exp2,exp3,exp4}.
Quantum wires hosting MF could also be engineered
in trapped ultracold fermionic atoms,
by employing optical Raman transitions to generate effective spin-orbit
coupling and Zeeman fields,
and using the proximity effect with a bulk molecular Bose-Einstein 
condensate\cite{demler}.
In these systems one  
can tune the parameters with high temporal precision and efficiently control
the amount of disorder. Thus they offer a unique possibility to
study in a very clean way the nonequilibrium dynamics of MF 
following time-dependent perturbations,
and in particular their formation when the system undergoes the
topological phase transition.

In this Letter we study the real-time dynamics of a 1D
quantum wire of finite length in contact with a superconductor and
in presence of SO coupling. The system  is driven out of 
equilibrium by varying the external magnetic field by means of different protocols.
The formation of MF is monitored by calculating the
evolution of the recently proposed  Majorana order 
parameter\cite{bena1}.
Remarkably we show that for sudden variations of the Hamiltonian  
the system relaxes towards a nonthermal state that does not exhibits
fully developed MF. Only for sufficiently slow protocols the 
system undergoes the topological phase transition, over a timescale which
increases with increasing length of the wire.
Finally we show that the manipulation of 
MF inside the T phase must obey precise temporal constraints in order 
to preserve the Majorana character. This is a crucial issue for the practical implementation
of topological quantum computing.

{\it Model and formalism.---}
The Hamiltonian of the coupled wire of length $\mathcal{L}$ is given 
by\cite{vonoppen,nota1}
\beq
H&=&\int_{0}^{\mathcal{L}} dx \Psi^{\dag}(x) ( 
-\partial_{x}^{2}/2m-\mu-i\a \partial_{x} \s_{y} 
+V_{z}\s_{z} )\Psi(x)
\nonumber \\
&+& \Delta (\psi_{\ua} (x) \psi_{\da} (x)+\mathrm{h.c.}), 
\quad
\Psi^{\dag}=(\psi^{\dag}_{\ua},\psi^{\dag}_{\da}),
\label{hamcont}
\eeq
where $\s_{i}$ are the Pauli matrices, $\psi_{\s}(x)^{(\dag)}$
annihilates (creates) an electron of mass $m$ and 
spin $\s$ at position $x$ in the wire, $\mu$ is the chemical 
potential, $\Delta$ is the strength of the proximity pairing field, 
and $\a$ and $V_{z}$ are the amplitudes of SO coupling and Zeeman 
field respectively.
This model displays a topologically trivial (TT) phase for $Q\equiv 
\m^{2}+\Delta^{2}-V_{z}^{2}>0$
and a T phase for $Q<0$, the phase transition occurring at 
$Q=0$\cite{sarma,vonoppen}.
The existence of MF in the T phase can be probed by considering the Majorana
polarization\cite{bena1}, defined as the anomalous local density of states 
\be
P(x,\omega)=-\frac{1}{\pi}\mathrm{Im}\int_{0}^{\infty}e^{i\w t}
\sum_{\s} 2i\bra   \{ \psi^{\dag}_{\s}(x,t) , \psi^{\dag}_{\s}(x,0) \} 
\ket ,
\label{pol}
\ee
where the average is taken over the Hamiltonian ground-state $| \Psi_{0}\ket$.
As discussed in Refs. \cite{bena1,bena2} a good order parameter to  
detect the T phase can be built from $P$ as
\be
\Phi=\int_{0}^{\mathcal{L}/2}dx 
P(x,0)=-\int_{\mathcal{L}/2}^{\mathcal{L}}dx P(x,0).
\ee
Indeed in the TT phase it holds  $\Phi=0$, 
while in the T phase
only the MF contribute to the 
order parameter yielding $\Phi=1$. 
The nonanalytic behavior of $\Phi$ at $Q=0$ indicates the occurrence 
of the phase transition, while values of $\Phi$ close to 1 
signal MF not fully developed. It is worth observing that for noninteracting fermions
[as in the case of Hamiltonian in Eq. (\ref{hamcont})] the polarization
takes the simpler form\cite{bena1} 
\be
P(x,\omega)= 2
 \sum_{n,\s} \delta (\omega -\epsilon_{n}) u^{(n)*}_{\s}(x)v^{(n)}_{\s}(x),
 \label{polnoni}
\ee
where $n$ denotes the 
$n$-th eigenstate of the system with eigenvalue $\epsilon_{n}$ and 
wavefunction $(u^{(n)}_{\ua},v^{(n)}_{\da},u^{(n)}_{\da},v^{(n)}_{\ua})$
expressed in the basis 
$(\psi^{\dag}_{\ua},\psi_{\da},\psi^{\dag}_{\da},\psi_{\ua})$.

The pair of (real) zero-energy  MF supported by the Hamiltonian $H$ in the T phase
are located at the two opposite edges of the wire\cite{vonoppen,sarma} and 
decay exponentially into the bulk.
The corresponding wavefunction 
must obey the constraint $u_{\s}=v_{\s}$ (that ensures the particle/antiparticle 
equivalence) and can be found by solving the 
auxiliary problem\cite{sarma}
\be
\left(
\begin{array}{cc}
-\partial_{x}^{2}/2m -\mu +V_{z} &  -\eta \Delta + \a \partial_{x}\\
 \eta \Delta - \a \partial_{x} & -\partial_{x}^{2}/2m -\mu -V_{z} 
\end{array}
\right)
\left(
\begin{array}{c}
    u_{\ua}(x) \\
    u_{\da}(x)
\end{array}
\right)
=0,
\ee
where $\eta =\pm 1$. The two choices of $\eta$ provide the MF located at 
the right and left 
boundary of the wire respectively.
In the following we assume $\eta=1$, since the calculation with 
$\eta=-1$ follows the same line of reasoning.
The MF wavefunction is then obtained by imposing the ansatz
\be
\left(
\begin{array}{c}
    u_{\ua}(x) \\
    u_{\da}(x)
\end{array}
\right)
=
\left(
\begin{array}{c}
    U_{\ua} \\
    U_{\da}
\end{array}
\right)
g(x),
\label{u1}
\ee
with $g(x)=e^{-x /\ell}$\cite{nota}.
The characteristic equation for $\ell$ has in principle 4 complex 
solutions for a given $\l$. However, in 
the T phase (i.e. for $Q<0$) and for any $\eta =\pm1$
only a single solution is real and ensures the normalizability of 
the wavefunction\cite{sarma}. In addition 
it is straightforward to verify that the
allowed $\ell$ for $\eta=\pm 1$ do coincide, as dictated by symmetry 
constraints. The coefficients 
$U_{\s}$ are easily obtained and read
\beq
U_{\ua}&=&\frac{\ell^{-1/2}}{\sqrt{1+\left(\frac{V_{z}-\frac{\ell^{-2}}{2m}-\mu}{\a/\ell+ \Delta}\right)^{2}}}, \nonumber \\
U_{\da}&=&\frac{\ell^{-1/2}}{\sqrt{1+\left(\frac{\a/\ell+ \Delta}{V_{z}-\frac{\ell^{-2}}{2m}-\mu}\right)^{2}}}.
\label{solution}
\eeq
From the above solution we obtain that the Majorana order parameter 
$\Phi=2\eta\ell \sum_{\s}U_{\s}^{2}$ 
is $1$ in the left half-wire and $-1$ in the
right half-wire, as it should be\cite{bena2}.

{\it Real-time evolution.---} 
We now study the real-time evolution of  $\Phi$
after the variation of the external magnetic field $V_{z}$ according 
to different protocols.
The explicit calculations are performed within the tight-binding 
version of the Hamiltonian in Eq. (\ref{hamcont}), which  
reads
\beq
H&=&\sum_{i=1}^{\mathcal{N}}
-\frac{v}{2}\left[
C^{\dag}_{i}C_{i+1} +\mathrm{h.c.} 
-(\mu-v)C^{\dag}_{i}C_{i} \right] \nonumber \\
 &-& \left.  \frac{\a}{2}(i C^{\dag}_{i}\s_{y}C_{i+1} +\mathrm{h.c.} ) 
 + V_{z} C^{\dag}_{i}\s_{z}C_{i} \right. \nonumber \\
 &+& \left. \Delta (c_{i\ua}c_{i\da}+\mathrm{h.c.}), \quad  
 C_{i}^{\dag}=(c_{i \ua}^{\dag},c^{\dag}_{i\da}) \right. ,
\label{Htb}
\eeq
with $\mathcal{N}=\mathcal{L}/a$ and $a$ the lattice spacing.
The mapping between the parameters of the continuum and lattice 
models is discussed in Ref. \cite{fisher}. In the following we 
express energies in units of the hopping $v$ and times in units of 
$1/v$. If $V_{z}\to V_{z}(t)$ the Hamiltonian becomes explicitly 
time-dependent $H \to H(t)$ and the dynamics of the system
is addressed by propagating the equilibrium 
ground-state $| \Psi_{0}\ket$ appearing in Eq. (\ref{pol}) according to 
$| \Psi_{0}(t)\ket=\mathcal{T}\mathrm{exp}[-i\int_{0}^{t} ds H(s)] | 
\Psi_{0}\ket$, where $\mathcal{T}$ is the time-ordering operator.
The problem is numerically solved by discretizing the time and calculating
the evolution of $| \Psi_{0} \ket $ within a time-stepping procedure
$|\Psi_{0}(t_{j}) \ket \approx \mathrm{exp}[-i H(t_{j})\delta t] | 
\Psi_{0}(t_{j-1}) \ket  $,
where $t_{j} = j \delta t$, $\delta t$ being a small time step and 
$j$ a positive integer\cite{perfetto1,perfetto2}.
We have considered ramp-like switching protocols of duration 
$\tau$ bringing the magnetic field from the initial value 
$V_{z}^{(i)}$ 
at $t=0$ [with corresponding Hamiltonian $H(0) \equiv H_{i}$] 
to the final value $V_{z}^{(f)}$ [with Hamiltonian 
$H(t>\tau)\equiv H_{f}$] which is 
maintained constant for $t>\tau$, i.e.  
$V_{z}(t)=\theta(\tau-t)[V^{(i)}_{z}+(V^{(f)}_{z}-V^{(i)}_{z})]t/\tau + 
\theta(t-\tau)V^{(f)}_{z}$.
The order parameter is then extracted by 
following Ref. \cite{bena1}. Since energy is not conserved during the 
temporal evolution, we calculate $\Phi(t)$ 
by integrating over $x$ the instantaneous polarization $P(x,E_{\mathrm{opt}}(t))$, 
where $E_{\mathrm{opt}}(t)$ is the minimum average of $H(t)$ over the
evolved  eigenstates of $H_{i}$,  denoted by $|\phi_{n}(t) \ket$\cite{nota3}. 
At every time the states corresponding to the mininum energy are 
always two,
with same value of $|E_{\mathrm{opt}}(t)|$, but with opposite signs due to the symmetry
of the problem.
Therefore the two optimal 
$|\phi_{\mathrm{opt}}(t) \ket$ 
[with wavefunctions $u^{(\mathrm{opt})}_{\s}(x,t),v^{(\mathrm{opt})}_{\s}(x,t)$ 
to be inserted in Eq. (\ref{polnoni})]
are the best approximation to the pair of MF that 
the system can provide at a given time. 

\begin{figure}[tbp]
\includegraphics[width=4.35cm]{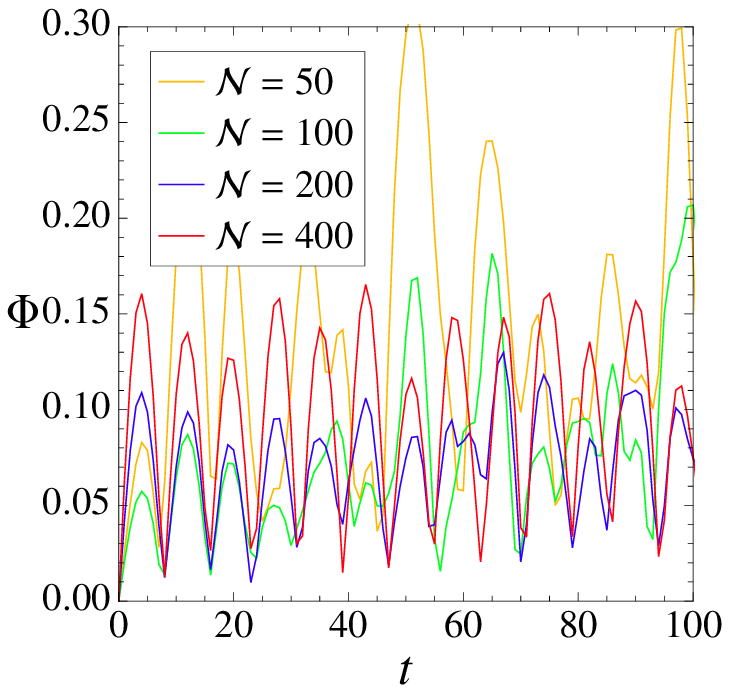}
\includegraphics[width=4.2cm]{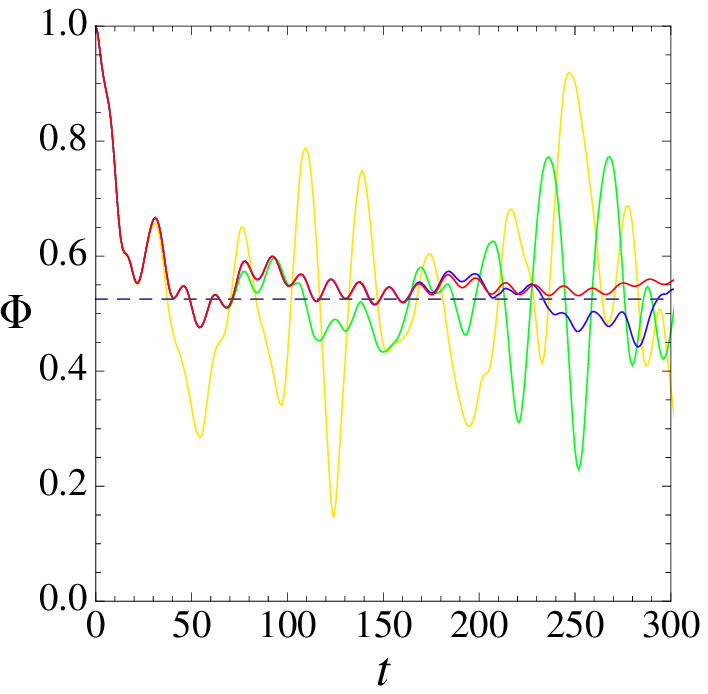}
\caption{$\Phi$ as a function of time for a
 $\mathrm{TT} \! \to \! \mathrm{T}$ quench with $V_{z}^{(i)}=10^{-6}$  
 and $V_{z}^{(f)}=0.2$ 
 (left panel), and for a $\mathrm{T} \! \to \! \mathrm{T}$ quench 
 with $V_{z}^{(i)}=0.2$ and $V_{z}^{(f)}=0.3$  
 (right panel). The rest of parameters are   $\mu =0$, $\Delta=0.1$, $\a =0.3$. 
 The dashed line in the right panel represents 
 $\Phi_{\mathrm{GGE}}$ calculated as in Eq. (\ref{phigge}).  
 $\Phi_{\mathrm{GGE}}$ has been obtained with $\mathcal{N}=400$, 
 but its value depends very weakly on 
 $\mathcal{N}$. }
\label{fig1}
\end{figure}

{\it Quench dynamics.---} 
If $\tau = 0^{+}$ we have a so-called sudden quench.
We have studied two relevant situations, namely $(i)$ the system initially   
in the TT phase and the value $V_{z}^{(f)}$ such that $Q<0$ for the 
final Hamiltonian ($\mathrm{TT} \! \to \! \mathrm{T}$ quench), and $(ii)$ $Q<0$ both for the initial and final 
systems ($\mathrm{T} \! \to \! \mathrm{T}$ quench).
The first case serves to understand how the 
MF are dynamically formed, while the second case is crucial to investigate 
whether the MF maintain their character after a manipulation of
the system parameters {\it inside} the T 
phase.
In Fig. \ref{fig1} we show the time evolution of the order parameter
in the two cases, for different values of the size $\mathcal{N}$.
Remarkably {\it if the system is initially in the $\mathrm{TT}$ phase the MF do not 
form} after the quench, and the order parameter remains close to zero 
for any $\mathcal{N}$, displaying temporal oscillations due to the 
finite size of the system (Fig. \ref{fig1} left panel). 
If instead the system is initially in the T phase, we see that the MF of $H_{i}$ 
are corrupted by the quench dynamics even if the condition $Q<0$ is 
maintained at all times, and $\Phi$ approaches finite values 
significantly smaller 
than 1 (Fig. \ref{fig1} right panel). Also in this case 
$\Phi(t)$ displays oscillations 
which, however, tend to disappear by increasing $\mathcal{N}$.    
We have verified that  similar results are also found by quenching 
other quantities like the strength of the pairing field, the SO coupling or the 
chemical potential (not show).
The above findings indicate that after a sudden quench the initial
ground-state $| \Psi_{0}\ket$ does not relax to the 
ground-state of the quenched 
Hamiltonian\cite{cazalilla,perfetto3,cazalilla2,perfetto4}, and, more in important, that no 
single-particle eigenstate of $H_{i}$ transforms into the MF of 
$H_{f}$. 
It has been conjectured\cite{rigol} that some of the properties of the nonthermal state
that develops after a sudden quench can be addressed within the so-called
generalized Gibbs ensemble (GGE).  The GGE permits to
compute a class of long-time averages of 
of integrable quenched systems by means of the 
special density matrix\cite{rigol}
\be
\rho_{\mathrm{GGE}}=\frac{1}{Z_{\mathrm{GGE}}}e^{-\sum_{n}\lambda_{n}\mathcal{I}_{n}},
\ee
where $Z_{\mathrm{GGE}}=\mathrm{Tr} 
[e^{-\sum_{n}\lambda_{n}\mathcal{I}_{n}}]$ and 
$\mathcal{I}_{n}$ are a set of integrals of motion of the quenched 
Hamiltonian. The weights $\l_{n}$ are fixed by the condition
$\mathrm{Tr}[ \rho_{\mathrm{GGE}}\mathcal{I}_{n}]=\bra 
\mathcal{I}_{n} \ket_{t=0}$. Thus one can argue that the average of an observable $O$ after 
the quench is obtained as $O_{\mathrm{GGE}} = \mathrm{Tr} 
[\rho_{\mathrm{GGE}} O]$. 
In the present case we choose $\mathcal{I}_{n}$ 
as the eigenmode-occupations of the  quenched
Hamiltonian, i.e.    
$\mathcal{I}_{n}=\gamma_{n}^{\dag}\gamma_{n}$ where 
$H_{f}=\sum_{n} \epsilon_{n} \gamma_{n}^{\dag}\gamma_{n}$.
As long as only the zero-energy states contribute to $\Phi$, the GGE 
is, in this case, spanned by the pair of MF.
Thus the order parameter 
calculated within the GGE takes an 
elegant analytic expression, being 
given by the square modulus of the overlap between
the (real) MF of the initial Hamiltonian
and those of the final Hamiltonian
\begin{figure}[tbp]
\includegraphics[width=7.35cm]{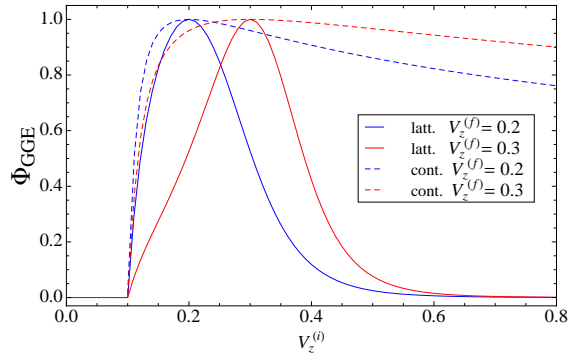}
\caption{$\Phi_{\mathrm{GGE}}$ for a $\mathrm{T} \! \to \! 
\mathrm{T}$ quench as a function of 
$V_{z}^{(i)}$ at fixed $V_{z}^{(f)}=0.2$ and $V_{z}^{(f)}=0.3$ for 
both the continuum and the lattice models. The
rest of parameters are the same as in Fig. \ref{fig1}.
We set $a=1$ and hence $m=1/v$\cite{fisher}. For the lattice model we 
computed numerically Eq. (\ref{phigge}) by diagonalizing $H$ with 
$\mathcal{N}=200$ (the result is however quite insensitive on the 
value of $\mathcal{N}$) while for the continuum model we used the 
analytic wavefunction in Eqs. (\ref{u1},\ref{solution}).}
\label{fig2}
\end{figure}
\be
\Phi_{\mathrm{GGE}}=\left|\sum_{\s}\int_{0}^{\mathcal{L}/2}dx \,  
u_{\s}^{(i)}(x) 
u_{\s}^{(f)}(x) \right| ^{2},
\label{phigge}
\ee
where
$i$ and $f$ denote the MF wavefunctions calculated with respect to  
$H_{i}$ and $H_{f}$ respectively.
As shown in Fig. \ref{fig1} the long-time average of $\Phi$ after 
the sudden quench is in very good agreement with 
the corresponding quantity calculated within the GGE.
For the $\mathrm{TT} \! \to \! \mathrm{T}$ quench
the GGE predicts $\Phi_{\mathrm{GGE}}=0$ since $H_{i}$ has no MF; accordingly the real-time 
simulations provide  values of $\Phi(t)$ close to 0, see left panel Fig. \ref{fig1}.
In the case of the  $\mathrm{T} \! \to \! \mathrm{T}$ quench the 
agreement is even more remarkable, see dashed line in right panel.  Indeed the oscillations 
of $\Phi(t)$ (sizable only for  $\mathcal{N}\lesssim 100$)  are
exactly centered around the value $\Phi_{\mathrm{GGE}}$. Thus we can 
infer that the GGE provides accurate predictions of the asymptotic quench 
dynamics of the Majorana order parameter. 

In Fig. \ref{fig2} we illustrate the robustness of the T phase after a $\mathrm{T} \! \to \! 
\mathrm{T}$ quench. The value  $\Phi_{\mathrm{GGE}}$ is plotted as a 
function of the strength of the initial magnetc field  
$V_{z}^{(i)}$, 
for a fixed value of the target $V_{z}^{(f)}$. We see that
if  $\Delta<V_{z}^{(i)}<V_{z}^{(f)}$ the T phase is readily corrupted,
even for small differences between the strengths of the initial and final magnetic 
fields. This happens in both the continuum and in the lattice models.
If $V_{z}^{(i)}>V_{z}^{(f)}$, instead, the continuum model predicts a 
significantly more stable T phase, since the system can afford larger 
changes in $V_{z}$ still maintaining values of $\Phi$ close to 1 (see 
Fig. \ref{fig2} dashed lines).

{\it Ramp protocols.---}
\begin{figure}[tbp]
\includegraphics[width=4.2cm]{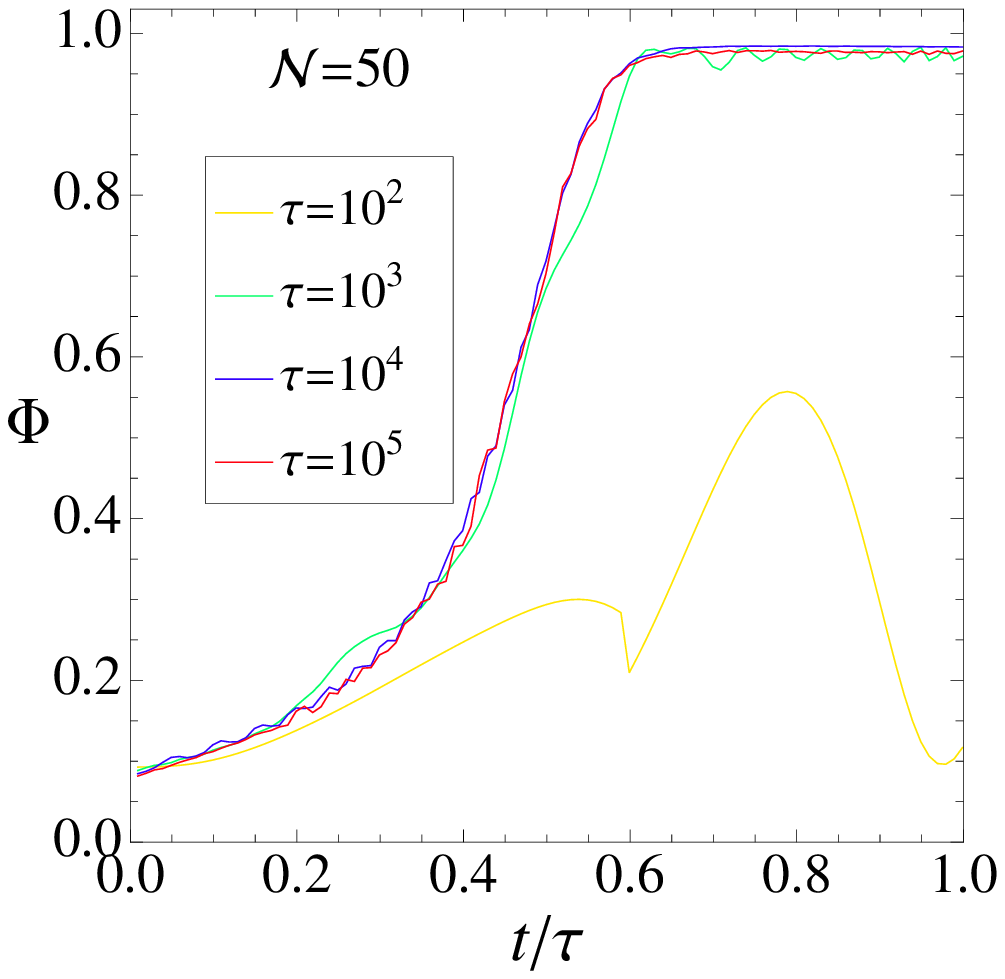}
\includegraphics[width=4.2cm]{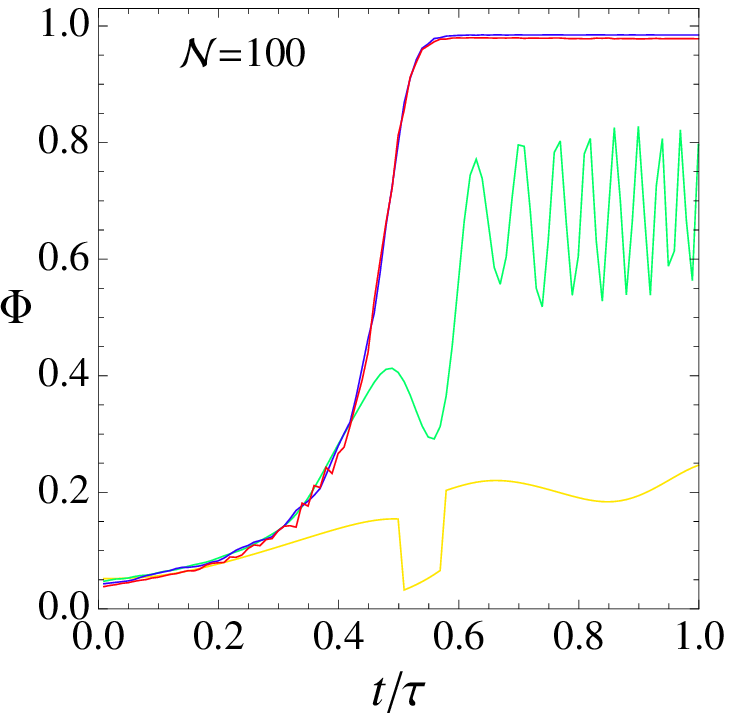}
\includegraphics[width=4.2cm]{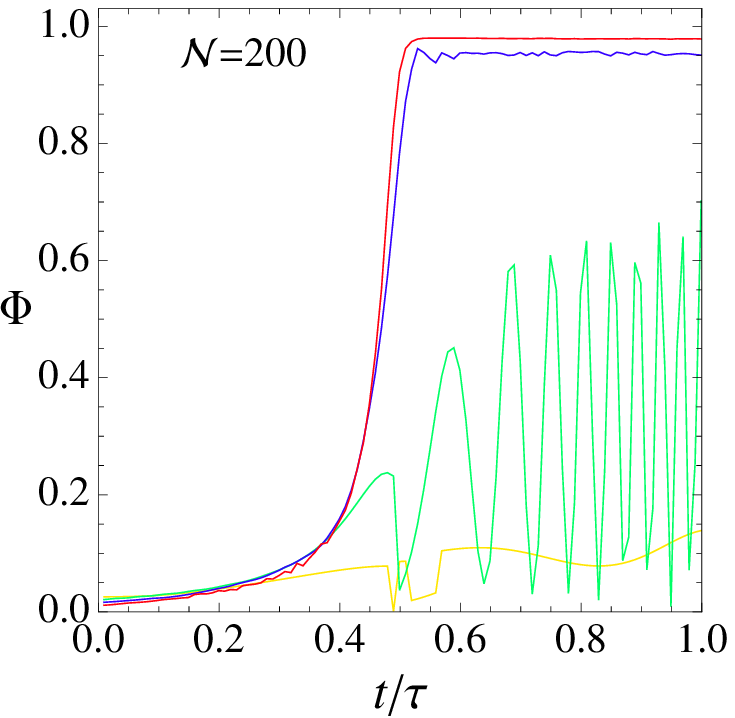}
\includegraphics[width=4.3cm]{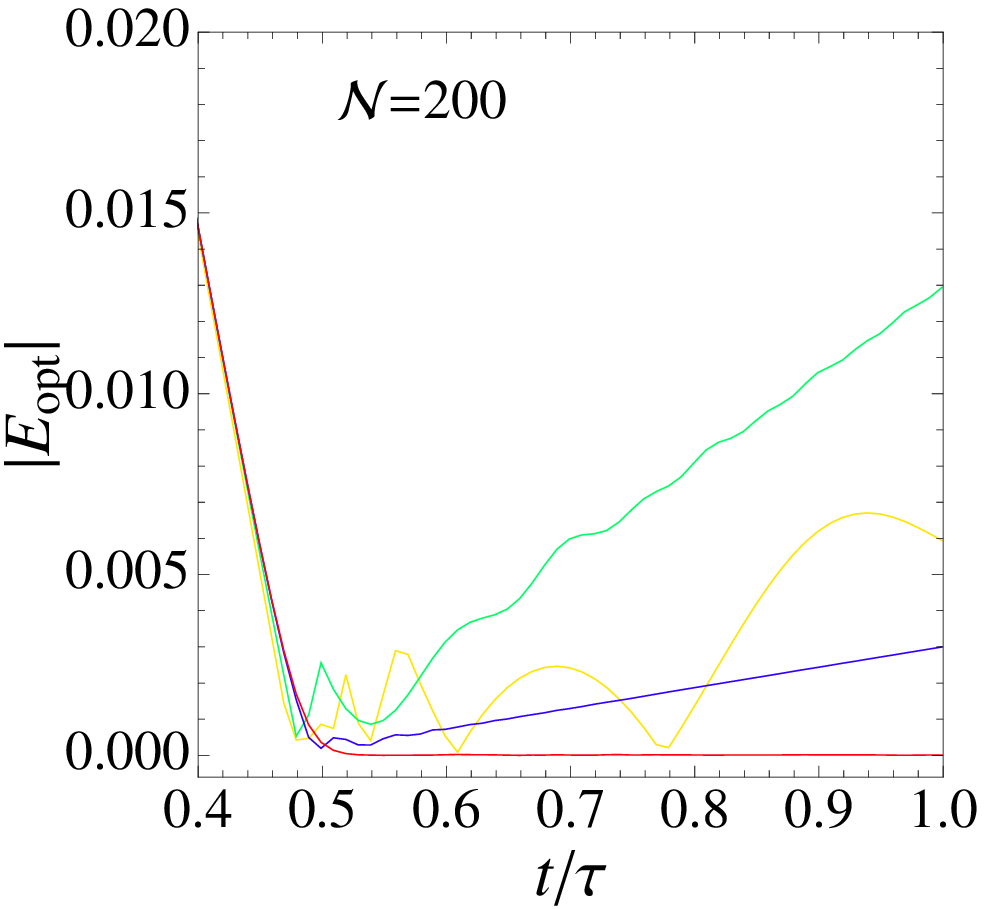}
\caption{
$\Phi$ and lowest energy $E_{\mathrm{opt}}$ as a function of time for $\mathrm{TT} \! \to \! 
\mathrm{T}$ transitions wit different ramp duration and wire 
length. The protocol parameters are $V_{z}^{(i)}=0.01$ and 
$V_{z}^{(f)}=0.3$ and the rest of parameters are the same as in Fig. 
\ref{fig1}. The discontinuities in $\Phi(t)$
observed for $\tau=10^{2}$ are due to dynamical level crossings in 
the set $\{ \varepsilon_{n}(t) \}$\cite{nota3}.
}
\label{fig3}
\end{figure}
We now investigate the possibility of generating/maintaining the T phase after an 
arbitrary change in the strength of the magnetic field 
by considering ramp protocols of finite duration $\tau$.  
In Fig. \ref{fig3} we plot $\Phi(t)$ for different $\mathrm{TT} \! \to \! 
\mathrm{T}$ transitions by varying $\tau$ and 
$\mathcal{N}$.
It is seen that for sufficiently slow ramps the topological phase 
transition takes place and MF fully develop, in agreement with the 
adiabatic hypothesis.
We mention, however, that the validity of the adiabatic theorem is 
not obvious in the present contest, since the final Hamiltonian is such that its square 
$H_{f}^{2}$ has a doubly degenerate 
ground-state, i.e. the pair of MF\cite{messiah}.
In the lower-right panel we also plot $E_{\mathrm{opt}}(t)$.
For $\mathcal{N}=200$ and $\tau =10^{5}$ the formation of the MF is 
clearly visible as $E_{\mathrm{opt}}$ approaches zero when $V_{z}(t)$ overcomes $\Delta$ (i.e. 
when $Q$ becomes negative).
Still, our results 
cleary indicate that the duration $\tau$ required to reach $\Phi=1$ increases
with increasing size. For instance $\tau \sim 10^{3}$ is required 
to create MF in a wire of length $\mathcal{N}=50$, whereas $\tau$ 
increases of about two orders of magnitude by enlarging the system to 
$\mathcal{N}=200$.
Interestingly this trend is reversed if one considers $\mathrm{T} \! \to \! 
\mathrm{T}$ transitions. In this case the larger is the size of the 
wire, the shorter is the ramp-time needed to maintain the Majorana 
character after the transformation. In Fig. \ref{fig4} we show the
duration $\bar{\tau}$ required for $\mathrm{T} \! \to \! \mathrm{T}$ transition
to reach a desired value $\bar{\Phi}$ of the order parameter (averaged over long times) close to 1.
It appears that $\bar{\tau}$ is an overall decreasing function of $\mathcal{N}$
(although with relatively small oscillations) that for the 
chosen parameters saturates to the finite value $\bar{\tau} \approx 80 $ for 
$\bar{\Phi}=0.95$ and $\bar{\tau} \approx 60 $ for 
$\bar{\Phi}=0.9$.
We have checked (not shown) that this qualitative behavior does not change by varying the features of 
the protocol, thus providing an explicit measure of adiabaticity 
for protocols aiming at preserving MF. 
\begin{figure}[tbp]
\includegraphics[width=7.35cm]{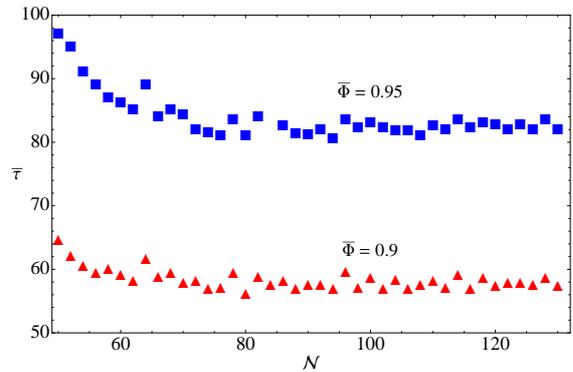}
\caption{Ramp-duration $\bar{\tau}$ required for a  $\mathrm{T} \! \to \! \mathrm{T}$ transition
to reach the asymptotic value $\bar{\Phi}=0.95$ (boxes) and    
$\bar{\Phi}=0.9$ (triangles) as a function of the length $\mathcal{N}$. The protocol parameters are  
$V_{z}^{(i)}=0.2$ and  $V_{z}^{(f)}=0.3$, and the rest of parameters 
are the same as in Fig. \ref{fig1}.
}
\label{fig4}
\end{figure}

{\it Conclusions.---}
We have studied the temporal evolution of MF in driven 1D quantum wires 
after the variation of the system parameters according to different protocols.
In the case of sudden quenches, thermalization breakdown is observed, and 
the long-time behavior of the Majorana order parameter $\Phi$ is well described within the 
GGE. Remarkably we find that the relaxed state does not display fully developed 
MF, no matter the initial Hamiltonian is: for a $\mathrm{TT} \! \to \! \mathrm{T}$
quench the MF are not generated and $\Phi$ remains close to zero, while for a $\mathrm{T} \! \to 
\! \mathrm{T}$ quench the MF initially present get readily corrupted 
as the system is driven out of equilibrium. 
The adiabatic theorem is, instead, recovered for extremely slow ramp protocols.
In the case of a slow $\mathrm{TT} \! \to \! \mathrm{T}$ ramp the system undergoes dynamically the topological phase 
transition within a timescale that increases by increasing the length 
of the wire. Finally we have provided an explicit 
estimate of adiabaticity to preserve MF during $\mathrm{T} \! \to \! 
\mathrm{T}$ protocols, by pointing out the existence of precise temporal constraints
relevant for the topological quantum computation.



\begin{thebibliography}{99}
\bibitem{nota}
We assume that $\mathcal{L}\gg \ell $ in order to calculate 
independently the  MBSs at the two edges, and also to simplify their 
normalization condition.
\end{thebibliography}
\end{document}